*Research Article*

# Nodes Localization in 3D Wireless Sensor Networks Based on Multidimensional Scaling Algorithm

**Biljana Risteska Stojkoska**

*Faculty of Computer Science and Engineering, Saints Cyril and Methodius University, 1000 Skopje, Macedonia*

Correspondence should be addressed to Biljana Risteska Stojkoska; biljana.stojkoska@finki.ukim.mk

In the recent years, there has been a huge advancement in wireless sensor computing technology. Today, wireless sensor network (WSN) has become a key technology for different types of smart environment. Nodes localization in WSN has arisen as a very challenging problem in the research community. Most of the applications for WSN are not useful without a priory known nodes positions. Adding GPS receivers to each node is an expensive solution and inapplicable for indoor environments. In this paper, we implemented and evaluated an algorithm based on multidimensional scaling (MDS) technique for three-dimensional (3D) nodes localization in WSN using improved heuristic method for distance calculation. Using extensive simulations we investigated our approach regarding various network parameters. We compared the results from the simulations with other approaches for 3D-WSN localization and showed that our approach outperforms other techniques in terms of accuracy.

## 1. Introduction

Wireless sensor network (WSN) represents a collection of wireless sensor nodes that coordinate with each other in order to perform a particular task. Sensor node is low-cost and low-power device that consists of three components: a sensing subsystem for data acquisition, a processing subsystem for local data processing, and a wireless communication subsystem for data transmission [1, 2]. Data are transferred from sensor nodes to the sink node (base station) through a multi-hop communication paradigm. Each sensor sends data to its closest neighbor responsible for retransmitting the packets.

Following the latest developments in computer and communication technologies, everyday objects are becoming smarter, as ubiquitous connectivity and modern sensors allow them to communicate with each other. The deployment of sensors and actuators everywhere around us adds a new dimension to the world of information and communication [3], which enables the creation of new and enriched services widely applied in different industrial and civilian application areas, including industrial process monitoring and control, machine health monitoring, environment and habitat monitoring, healthcare applications, and traffic control [4, 5].

A fundamental problem in wireless sensor networks is localization, that is, the determination of the geographical locations of sensors. The most straightforward solution to the localization problem is to apply global positioning system (GPS) to each node. But it is not an attractive solution because of cost, size, and power constraints [6–8]. Thus, an effective localization algorithm should employ all the available information from the nodes to compute the positions.

Localization is a challenge when dealing with wireless sensor nodes and a problem which has been studied for many years. Many different techniques have been proposed for solving this problem, but most of them consider only two-dimensional (2D) network. Hence, localization issue in three dimensions remains a challenging problem in the research community.

In this paper, we analyze the well-known MDS-MAP algorithm for nodes localization in three-dimensional WSN. Since MDS-MAP is one of the algorithms that produce the best results and is considered as referent algorithm by many researchers for localization in two-dimensional networks, we were motivated to implement it and investigate its performances in three-dimensional networks. Additionally, we propose a new algorithm based on MDS, which uses a heuristic approach for distance matrix calculation, thus we improved the accuracy of MDS-MAP. Henceforth, we would refer to our algorithm as IMDS.

The rest of this paper is organized as follows. In the second section, the relevant work related to the present 3D localization techniques is discussed. The third section refers to multidimensional scaling as a technique for nodes localization in three-dimensional WSN and covers its mathematical background. The fourth section elaborates our IMDS that uses a heuristic approach for distance matrix calculation. Section five and section six present the results provided from the simulations for 3D network and for 3D surface networks, respectively. Finally, we conclude this paper in section seven.

## 2. Related Work

Techniques for WSN localization can be basically divided into two categories: range-based and range-free methods. The range-based techniques are considered more accurate and most of the algorithms for localization belong to this category [6, 9]. They use the distance between the nodes in the network [10, 11]. RSSI (receive signal strength indicator) is the most common technique used for distance estimation. RSSI utilizes small resources without the need for extra hardware. RSSI measures the power of the received radio signal to calculate the distance between two nodes that are in transmission range of each other [12]. Other techniques (time of arrival: ToA, time difference of arrival: TDoA, etc.) for distance measurement translate propagation time into distance [11]. This can be done if signal propagation speed is known in advance. These techniques can be used with acoustic, infrared, and ultrasound signals.

Many research groups have investigated different techniques for nodes localization in WSN, but most of the proposed techniques consider only two-dimensional WSN localization. A few researchers within the last years have tried to focus on three-dimensional localization.

One of the first proposed techniques for 3D localization is Landscape-3D [13]. In the first phase, location-unaware nodes measure a set of distances to mobile location assistants (LAs) using RSSI. In the next phase, nodes use unscented Kalman filter to estimate their own position. Very similar approach is proposed in [14]. RSSI is used for distance measurements while particle filter is used for nodes positioning.

Although these methods are independent of networks density and networks topology, the major drawback is its dependence on mobile devices that might not be available under some deployments (e.g. hostile environments).

In [15], the authors propose cluster-based approach named CBLALS. In each cluster, the intercluster range measurement errors are corrected using triangle principle. The evaluation of CBLALS with respect to other approaches shows that CBLALS has much better positioning accuracy.

A novel centroid localization method that significantly improves the basic centroid localization algorithm is presented in [16]. Each unknown node randomly selects four anchor nodes (nodes whose coordinates are known in advance) in range to form a series of tetrahedrons used to calculate its own position using novel centroid method.

## 3. Multidimensional Scaling for Localization in 3D-WSN

Multidimensional scaling (MDS) is a set of techniques used for reducing the dimensionality of the data (objects). MDS visualizes the results in order to show hidden structures in the data [17]. MDS algorithm uses the distances between each pair of object as an input and generates 2D-points or 3D-points as an output.

In WSN, MDS has an ability to reconstruct the relative map of the network even without anchor nodes [18, 19]. The first and most explored is MDS-MAP [19], which is very accurate for density networks. More computationally dependent approaches are introduced in [20]. MDS-MAP(P) [20], as a modification of MDS-MAP, computes local maps at each node in the network and then integrates them into a global map. Different approaches based on MDS-MAP are introduced in [21, 22]. In cluster-based MDS [22], the network is divided into clusters responsible for partial localization using MDS-MAP. This approach, evaluated for irregular network topologies, shows better performances than MDS-MAP in terms of accuracy and better performances than MDS-MAP(P) in terms of computational complexity. There are dozens of algorithms in the literature based on MDS, but most of them are developed and investigated for 2D WSN. In our previous work, MDS-MAP was implemented and evaluated for 3D WSN [23]. In this paper, we extended our work from [23], and propose a modification of MDS that should improve the accuracy. In this part, we will explain in-depth multidimensional scaling algorithm for three-dimensional networks.

*3.1. Mathematical Background of MDS-MAP in Three Dimensional Space.* Consider a WSN with $n$ nodes in a three dimensional space. Let $S_{n\times 3}$ be an unknown matrix where each row represents the coordinates of $i$-th point (node) along three dimensions. Let $\Delta^{(2)}(S)$ or $\Delta^{(2)}$ represents the matrix of squared Euclidean distances between nodes $i$ and $j$ $(i = 1\dots n,\ j = 1\dots n)$. The $\Delta^{(2)}$matrix can be calculated since we assume that nodes have mechanisms to estimate the distance between each other:

$$\Delta^{(2)}(S) = \Delta^{(2)} = \begin{bmatrix} 0 & \Delta^{(2)}{}_{12} & \Delta^{(2)}{}_{13} & \cdots & \Delta^{(2)}{}_{1n} \\ \Delta^{(2)}{}_{21} & 0 & \Delta^{(2)}{}_{23} & \cdots & \Delta^{(2)}{}_{2n} \\ \cdots & & & & \\ \Delta^{(2)}{}_{n1} & \Delta^{(2)}{}_{n2} & \Delta^{(2)}{}_{n3} & \cdots & 0 \end{bmatrix}, \tag{1}$$

where

$$\Delta^{(2)}{}_{ij}(S) = \Delta^{(2)}{}_{ij} = \sum_{a=1}^{3} \left(s_{ia} - s_{ja}\right)^2. \tag{2}$$

It can be rewritten as

$$\Delta^{(2)}{}_{ij} = \sum_{a=1}^{3} \left(s_{ia}{}^2 + s_{ja}{}^2 - 2 s_{ia} s_{ja}\right), \tag{3}$$

or

$$\Delta^{(2)} = c1' + 1c' - 2SS', \tag{4}$$

where 1 is an $n \times 1$ vector of ones and $c$ is a vector consisting the diagonal elements of the scalar product matrix; that is, $c = \sum_{a=1}^{3} s_{ia}{}^{2}$.

Multiplying both sides of (4) by centering matrix $T$

$$T = I - n^{-1}11', \tag{5}$$

where $I$ is the identity matrix, we get

$$\begin{aligned} T\Delta^{(2)}T &= T\left(c1' + 1c' - 2SS'\right)T, \\ T\Delta^{(2)}T &= Tc1'T + T1c'T - T2SS'T, \\ T\Delta^{(2)}T &= Tc1'T + T1c'T - T(2B)T, \\ T\Delta^{(2)}T &= -T(2B)T, \\ B &= -\frac{1}{2}T\Delta^{(2)}T. \end{aligned} \tag{6}$$

Since $B$ is symmetric it can be decomposed:

$$B = Q\Lambda Q' = \left(Q\Lambda^{1/2}\right)\left(Q\Lambda^{1/2}\right)' = SS', \tag{7}$$

where $Q$ is orthonormal and $\Lambda$ is a diagonal matrix

$$\Longrightarrow S = Q\Lambda^{1/2}. \tag{8}$$

The recovered matrix $S$ obtained from (8) represents a relative map and has to be rotated, as it has a different coordinate system.

MDS-MAP for 3D WSN consists of 3 steps as follows.

(1) Calculate the shortest distances between every pair of nodes (using either Dijkstra's or Floyd's all pairs shortest path algorithm). This is the distance matrix that serves as an input to the multidimensional scaling in step 2.

(2) Apply classical multidimensional scaling to the distance matrix. The first 3 largest eigenvalues and eigenvectors give a relative map with relative location for each node.

(3) Transform the relative map into absolute map using sufficient number of anchor nodes (at least 4). This process usually includes translation, rotation, and reflection.

*3.2. Finding Optimal Rotation and Translation between Corresponding 3D Nodes.* Generating an absolute map (step 3) of the WSN requires anchor nodes. At least four sensors' physical positions are needed in order to identify the physical positions of the remaining nodes.

Let $P = \{p_1, p_2, \ldots, p_N\}$ and $Q = \{q_1, q_2, \ldots, q_N\}$ be two sets of corresponding nodes, where $N$ is the number of anchor nodes in the WSN. We need to find a transformation that optimally aligns the two sets in terms of least square errors, that is, to minimize the sum of squares of the errors between estimated positions of the anchors from MDS map and their true positions. We seek a rotation matrix $R$ and a translation vector $t$ such that

$$(R, t) = \arg\min_{R,t} \sum_{i=1}^{N} \left\|\left(Rp_i + t\right) - q_i\right\|^2. \tag{9}$$

This transformation is also known as Euclidean or rigid transformation, because it preserves the shape and the size.

There are many algorithms proposed in the literature that compute a rigid 3D transformation. Among them, the most explored are those based on:

(i) singular value decomposition (SVD),

(ii) unit quaternion (UQ),

(iii) dual quaternion (DQ),

(iv) orthonormal matrices (OM).

A comparison of these four methods can be found in [24]. It is shown in [25], that the results of all these methods are similar in most cases and the difference in accuracy is almost insignificant, but the SVD is the most stable.

Finding the optimal rigid transformation with SVD can be broken down into the following steps.

(i) Compute the weighted centroids of both point sets

$$\overline{p} = \frac{1}{N}\sum_{i=1}^{N} p_i, \qquad \overline{q} = \frac{1}{N}\sum_{i=1}^{N} q_i. \tag{10}$$

(ii) Compute the centered vectors

$$p_i{}' := p_i - \overline{p}, \qquad q_i{}' := q_i - \overline{q}, \quad i = 1, \ldots, N. \tag{11}$$

(iii) Compute the $3 \times 3$ covariance matrix

$$H = P'Q'^{T}, \tag{12}$$

where $P'$ and $Q'$ are the $3 \times N$ matrices that have $p_i{}'$ and $q_i{}'$ as their columns, respectively.

(iv) Compute the singular value decomposition

$$H = U\Sigma V^{T}. \tag{13}$$

The rotation we are looking for is

$$R = VU^{T}. \tag{14}$$

(v) Compute the optimal translation as

$$t = \overline{q} - R\overline{p}. \tag{15}$$

*3.3. Time Complexity of MDS-MAP for 3D-WSN.* In step 1, distance matrix construction using Dijkstra's or Floyd's algorithm requires $O(n^3)$, where $n$ is the number of nodes in the network [26]. In step 2, applying MDS to the distance matrix has complexity of $O(n^3)$ due to singular value decomposition. In step 3, the relative map is transformed through linear transformations. Computing the rigid transformation takes $O(N)$ time for computing $P$ and $Q$, while computing SVD takes only $O(3^3)$ time (since the covariance matrix $H$ has dimension $3 \times 3$). Applying the transformation (rotation and translation) to the whole relative map takes $O(n - N)$ time, where $N$ is the number of anchors ($N \ll n$).

## 4. MDS-Based Algorithm with Heuristic Approach for Distance Matrix Calculation (IMDS)

The main drawback of MDS-MAP is the way it calculates the distance matrix. Using Dijkstra (or Floyd's) all pairs shortest path algorithm to estimate the distances between nonneighboring nodes in the network gives incorrect distance matrix. Dijkstra distance between two nodes usually correlates with the Euclidean distance but always calculates the longest possible distance.

To reduce the error present in distance matrix, we introduce a heuristic approach (HA) to estimate the distances between nonneighboring nodes.

In this section, we will explain in detail our algorithm based on multidimensional scaling algorithm for nodes localization in WSN that uses HA for distance matrix calculation (IMDS).

*4.1. Dijkstra Algorithm for Distance Matrix Calculation.* MDS is very accurate technique for dimensionality reduction. If the correct distance matrix is given as input, MDS algorithm will reconstruct the map of the network without error. But, calculating distance matrix for networks where only distances between neighboring nodes are known is not a trivial task. This problem in MDS-MAP is solved by applying Dijkstra's (or Floyd's) all pairs shortest path algorithm. Dijkstra's algorithm is a graph search algorithm that solves the single-source the shortest path problem. In WSN localization problem, the sensor network is represented as a graph with nonnegative edge path costs, while the real, Eucledian distance between two nonneighboring nodes is replaced with the distance calculated using Dijkstra algorithm. But the assumption that Dijkstra distance between two nodes correlates with their Euclidean distance is hardly true. This approximation produces an error; that is, the positions obtained as MDS output usually differ from the correct positions. The difference between the real and the predicted positions is known as estimation error. The error is bigger when the nodes are in multihop communication range, which is a common case in obstructed environments. It is usually caused by the presence of obstacles or terrain irregularities that can obstruct the line of sight between nodes or cause signal reflections. Figure 1 shows two examples when Dijkstra algorithm will calculate much larger distance between nonneighboring nodes. Left side of the picture shows an example of two nodes $A$ and $B$ that are far from each other. The distance between $A$ and $B$ will be calculated as $AB = a + b + c + d$, which is much longer then the real Euclidian distance. This scenario is present when the network is deployed on vast regions where the radio range of the nodes is short compared with the length of the region. On the right side of Figure 1, there is an example where two nodes ($A$ and $B$) cannot communicate directly although they are very close to each other. The reason for this is the presence of obstacle that obstructs the line of sight. In this scenario, Dijkstra algorithm is completely inapplicable as it calculates a few times longer distance.

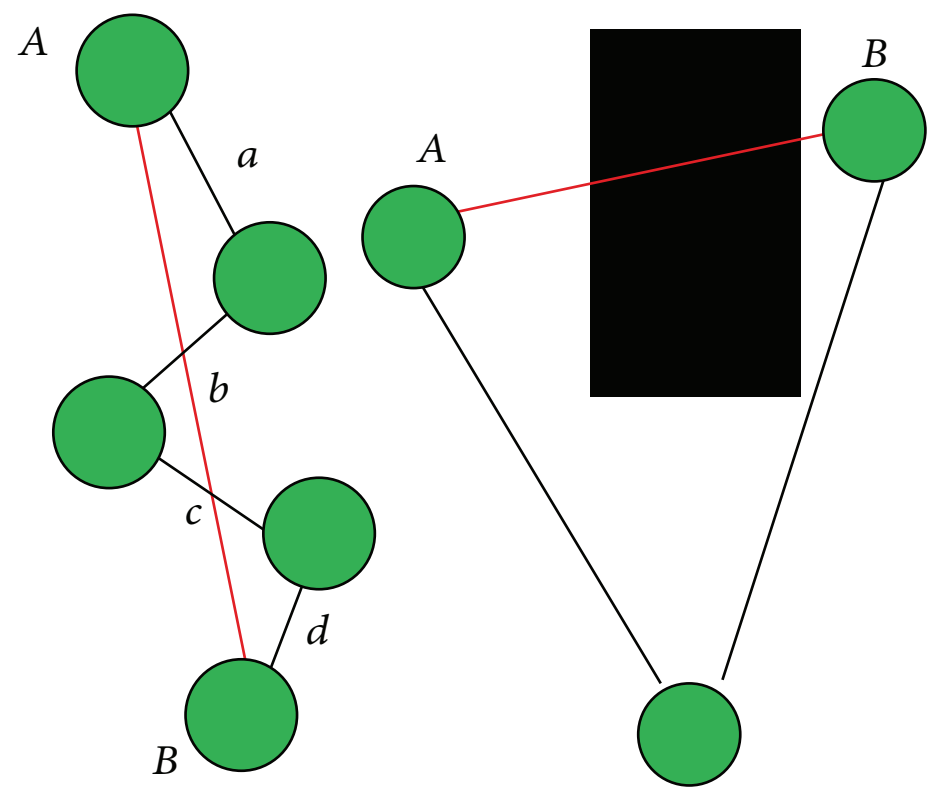

Figure 1: Distance calculation using Dijkstra algorithm.

*4.2. Heuristic Approach for Distance Matrix Calculation (HA).* As it can be seen from the two examples presented in Figure 1, the distance calculated using Dijkstra algorithm always increases the real distance. In order to reduce this distance, in this paper, we propose an alternative heuristic approach. By reducing the distance matrix error, we intend to reduce the overall estimation error.

Consider there are three nodes in a network, $A$, $B$, and $C$ (Figure 2), with known distances between nodes $A$ and $B$ ($d_1 = AB$) and between nodes $B$ and $C$ ($d_2 = BC$). Since distance matrix requires the distances between every pair of nodes in the network, the distance between nodes $A$ and $C$ has to be obtained. We will refer to this distance as $a$.

If maximum radio range of the nodes in the network is $R$, then we know for sure that node $C$ can lay anywhere on the curve $C_1C_2$. If Dijkstra's algorithm is used for this purpose, it will calculate the distance $a$ as $a = AB + BC$, which is the longest possible theoretical distance between nodes $A$ and $C$. More precisely, $C$ will lay exactly on $C_2$. On the other hand, if we calculate the shortest possible theoretical distance between nodes $A$ and $C$, it will be very close to $R$. We can conclude that

$$R < a \leq d_1 + d_2. \tag{16}$$

To minimize the possible error, we purpose a heuristic solution that assumes that the node $C$ lies exactly in the

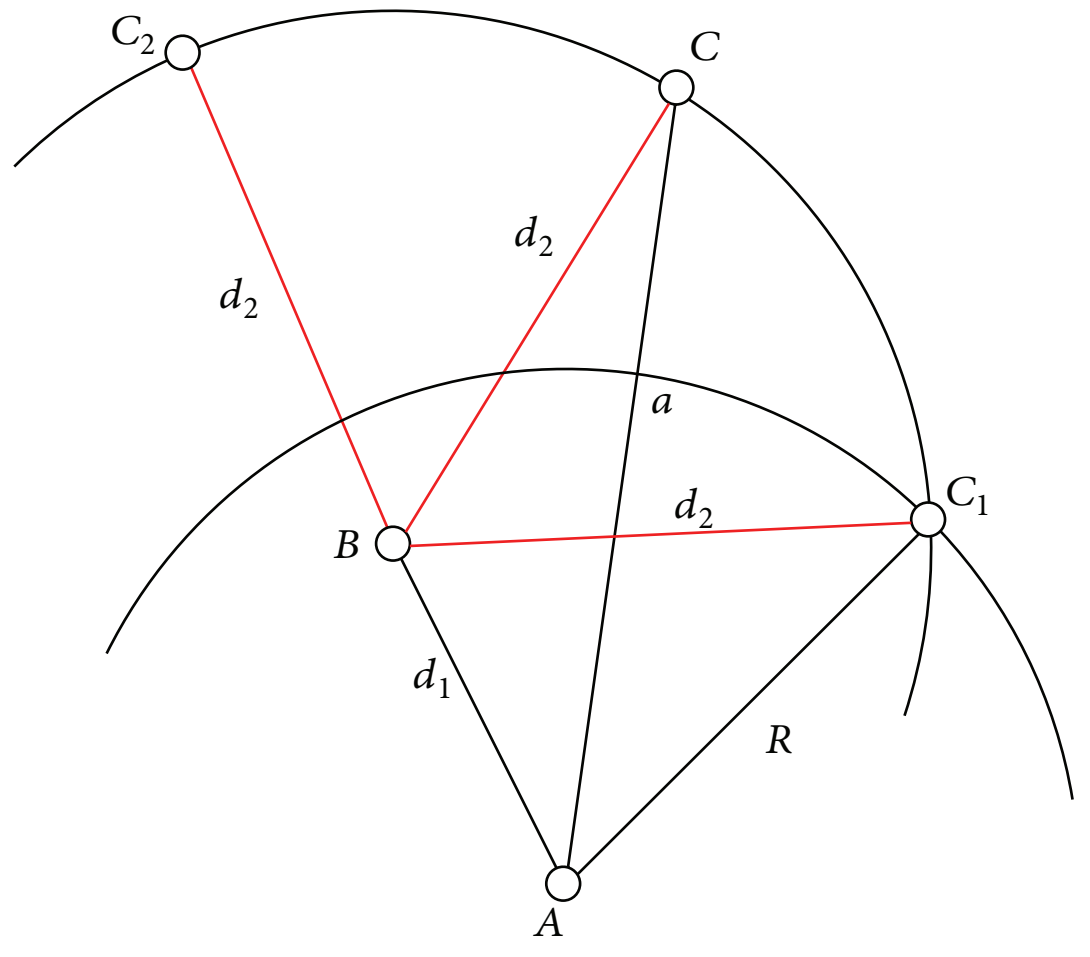

FIGURE 2: Distance calculation using Heuristic Approach (HA).

middle of the curve $C_1C_2$. Hence, the distance $a = AC$ can be calculated using cosine formula as

$$a^2 = {d_1}^2 + {d_2}^2 - 2 \cdot d_1 \cdot d_2 \cdot \cos(\sphericalangle ABC). \quad (17)$$

In order to calculate the distance $a$, first, we need to find the angle using cosine formula:

$$\sphericalangle ABC = \sphericalangle ABC_1 + \sphericalangle C_1BC. \quad (18)$$

The angle $\sphericalangle ABC_1$ can be calculated again with the cosine formula:

$$\sphericalangle ABC_1 = \arccos\left(\frac{{d_1}^2 + {d_2}^2 - R^2}{2 \cdot d_1 \cdot d_2}\right). \quad (19)$$

Since

$$\begin{aligned} \sphericalangle C_1BC &= \sphericalangle CBC_2, \\ \sphericalangle C_1BC &= \frac{1}{2}\sphericalangle C_1BC_2, \\ \sphericalangle C_1BC &= \frac{1}{2}(\pi - \sphericalangle ABC_1), \\ \sphericalangle ABC &= \sphericalangle ABC_1 + \frac{1}{2}(\pi - \sphericalangle ABC_1), \\ \sphericalangle ABC &= \frac{\pi}{2} + \frac{1}{2}\sphericalangle ABC_1. \end{aligned} \quad (20)$$

Finally,

$$\begin{aligned} a^2 &= {d_1}^2 + {d_2}^2 - 2 \cdot d_1 \cdot d_2 \cdot \cos(\sphericalangle ABC) \\ &= {d_1}^2 + {d_2}^2 - 2 \cdot d_1 \cdot d_2 \cdot \cos\left(\frac{\pi}{2} + \frac{1}{2}\sphericalangle ABC_1\right) \\ &= {d_1}^2 + {d_2}^2 + 2 \cdot d_1 \cdot d_2 \cdot \sin\left(\frac{1}{2}\sphericalangle ABC_1\right), \end{aligned} \quad (21)$$

where

$$ABC_1 = \arccos\left(\frac{{d_1}^2 + {d_2}^2 - R^2}{2 \cdot d_1 \cdot d_2}\right). \quad (22)$$

*4.3. Comparison of HA and Dijkstra.* In order to evaluate our heuristic approach (HA) for distance matrix calculation, we perform a simulation. We randomly placed 100 nodes in a square and calculated the distances between the nodes using both HA and Dijkstra algorithm. Then we compared the differences between the real distances and the distances obtained using the above-mentioned algorithms. The difference is normalized with radio range $R$. The results from the simulation are presented in Figure 3.

As can be seen from the figure, HA performs better than Dijkstra for all connectivity levels. This is expected knowing that Dijkstra calculates the longest possible distance, while HA tends to shorten this distance.

HA performs much better than Dijkstra especially for large range error $e_r$. This is rather important characteristic of HA since range measurement in the real applications is prone to error.

When RSSI is used for distance signalization, the range error measurement is at least 10% $R$. The results presented in [27] show average range error measurement between 5% R and 30% $R$ for longer radio range $R$. Similar research conducted in [28, 29] that investigate RSSI reported average measurement error around 20% $R$.

*4.4. IMDS Algorithm.* IMDS for 3D WSN consists of the following 3 steps.

(1) Calculate the shortest distances between every pair of nodes using heuristic approach. This is the distance matrix that serves as an input to the multidimensional scaling in step 2.

(2) Apply classical multidimensional scaling to the distance matrix. The first 3 largest eigenvalues and eigenvectors give a relative map with relative location for each node.

(3) Find the optimal rigid transformation with SVD and transform the relative map into absolute map using sufficient number of anchor nodes (at least 4).

IMDS preserves the time complexity of MDS-MAP algorithm.

## 5. Results and Discussion

We assume a typical sensor network composed of hundreds of sensor nodes deployed uniformly across a three-dimensional monitored area. Each sensor is equipped with an omnidirectional antenna and only nodes within certain radio range $R$ can communicate with each other. We made the following assumptions.

(i) Nodes are static and unaware of their location.

(ii) There is a path between every pair of nodes.

(iii) Nodes deployed in close proximity to each other exchange messages.

(iv) Each node uses RSSI (or any other) method for distance estimation.

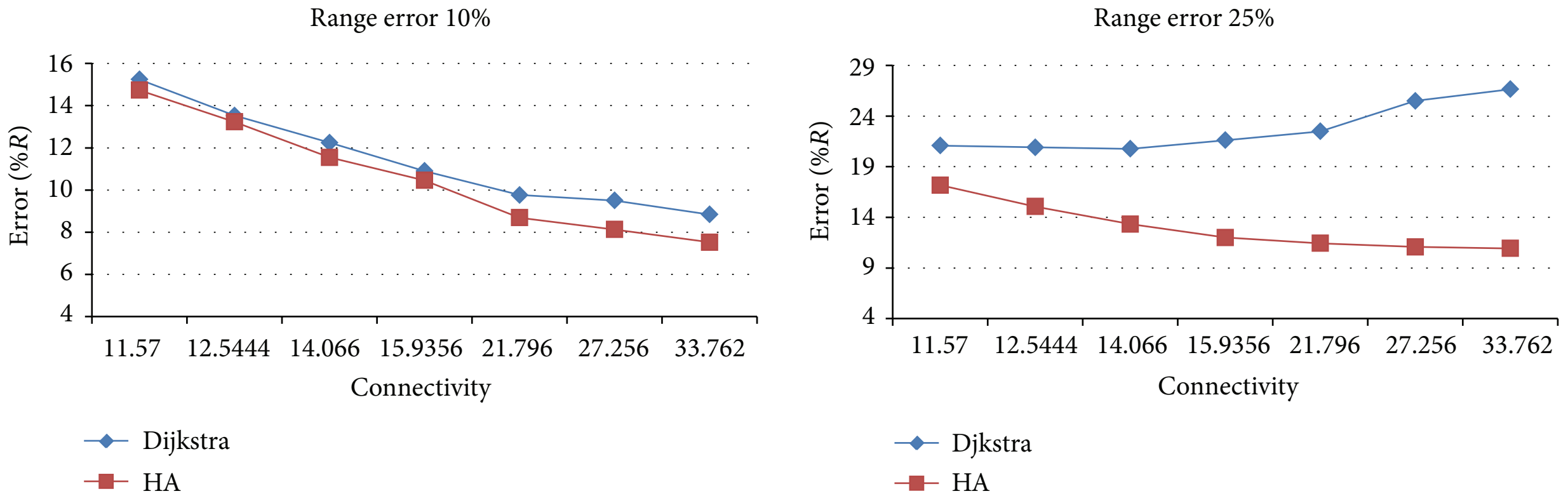

FIGURE 3: Distance error for HA and Dijkstra.

We simulated both IMDS and MDS-MAP on random network topology with Matlab. 100 nodes were uniformly placed in a cubic area (100 r × 100 r × 100 r, where r is a unit length distance). Our work was mainly focused on network properties like number of anchors, average connectivity, and range error. We consider the following.

(i) Different network topologies:

  (a) random deployment for 3D-WSN (100 nodes),
  (b) grid deployment for 3D-WSN (125 nodes).

(ii) Different number of anchors (4, 6, 10, and 15) for absolute map construction. In our experiments we use SVD method for 3D rigid transformation. In our simulation, the anchors were selected randomly.

(iii) Different radio ranges ($R$) which lead to different average connectivity (average number of neighbors).

(iv) Different radio range error $e_r$ (from 0 to 30% of $R$ with step 5% of $R$).

Thus 280 different networks were simulated (2 × 4 × 5 × 7) and each node location was discovered with both MDS-MAP and IMDS technique. The connectivity parameter and the estimation error for each scenario represent an average over 30 trials. The average estimation error is normalized by the radio range $R$:

$$\text{Error} = \frac{\sum_{i=1}^{(n-N)} \text{distance}\left(\text{pos}_i^{(\text{estimated})}, \text{pos}_i^{(\text{true})}\right)}{(n-N)\cdot R} \cdot 100\%, \tag{23}$$

where $n$ is the number of nodes in the network, $N$ is the number of anchor nodes, $\text{pos}_i^{(\text{estimated})}$ is the estimated location, and $\text{pos}_i^{(\text{true})}$ is the true location of the $i$-th node.

*5.1. Comparison of MDS-MAP and IMDS for 3D WSN.* Figure 4 shows an example of typical 3D network with 100 nodes randomly deployed ($R = 35$ r and an average connectivity of 11.6). Blue lines represent the distance between the absolute and the estimated position when using IMDS and MDS-MAP algorithm, respectively. The estimation error is larger if the lines are longer. The absolute map is achieved using 10 anchors (red circles). As can be seen from the figure, IMDS performs better than MDS-MAP.

Figure 5 shows the results of MDS-MAP and IMDS for random topology with range error 10% of $R$ and 15% of $R$ when using 10 anchors. As can be seen from the figure, in both cases IMDS performs smaller estimation error than MDS-MAP for all connectivity levels.

We must note that range error $e_r$ has great impact on estimation accuracy. When range error $e_r$ is small, MDS-MAP and IMDS perform almost the same, with IMDS being slightly better. As $e_r$ increases, MDS-MAP rapidly deteriorates, while IMDS is pretty much stable. Considering the fact that range error is inevitable phenomena in WSN, we can conclude that IMDS is better option when choosing localization algorithm for 3D environments.

As expected, using more anchor nodes gives slightly smaller estimation error. Number of anchors affects the results when the connectivity level is low. For high connectivity levels, there is no evident improvement (Figure 6).

*5.2. Comparison of MDS-MAP and IMDS with Other Approaches for 3D Localization.* To evaluate IMDS more convincingly, the simulation experiments were conducted for CBLALS method [15] and for the novel centroid algorithm [16]. The comparison with [15] shows that both IMDS and MDS-MAP perform better than CBLALS in terms of localization accuracy (Figure 7(a)). It is assumed that the density of the anchors is 10%, and the connectivity of the networks is 10 ~ 15. As can be seen from the results, MDS-MAP is much more robust to range error than CBLALS, especially for large range errors. The result from the comparison of IMDS and MAP-MAP with the novel centroid algorithm [16] is shown in (Figure 7(b)). For all different radio range $R$, both IMDS and MDS-MAP perform better than novel centroid algorithm. Here the density of the anchors is 20%. The radio range error for this simulation is $e_r = 0\%\ R$.

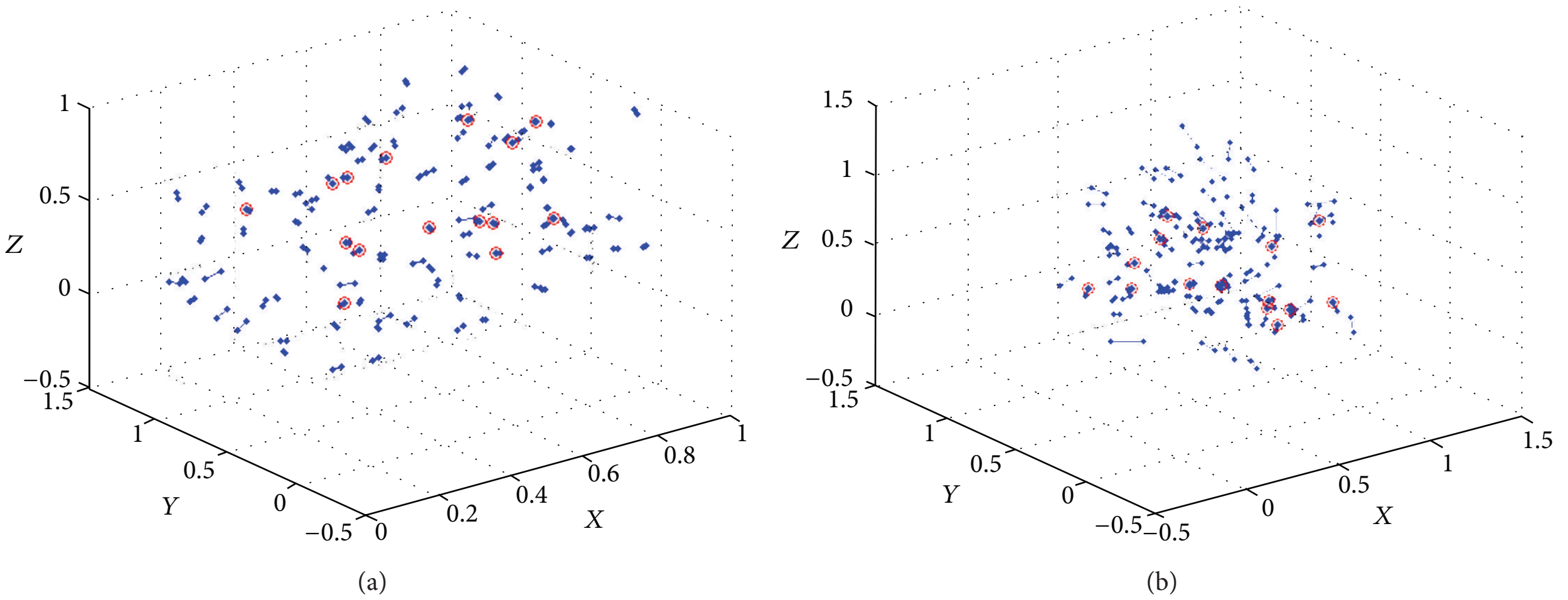

Figure 4: Estimation error for IMDS (a) and MDS-MAP (b).

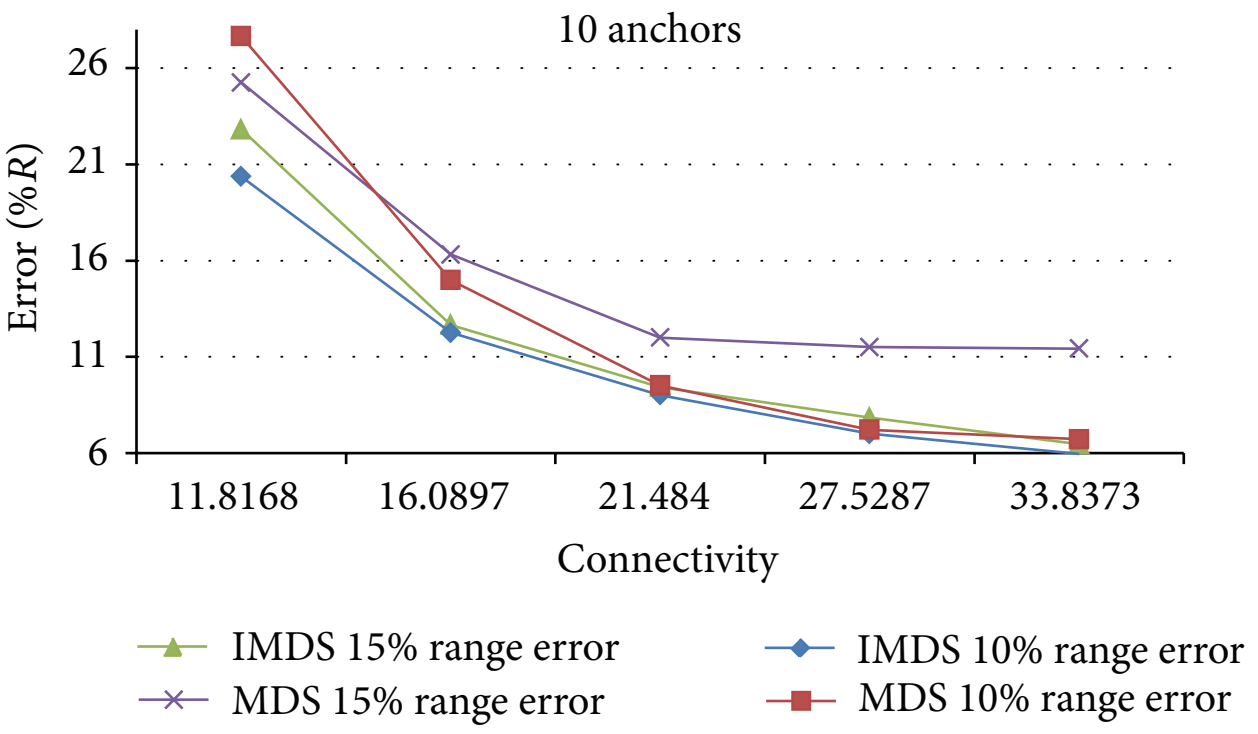

Figure 5: Estimation error for IMDS and MDS-MAP.

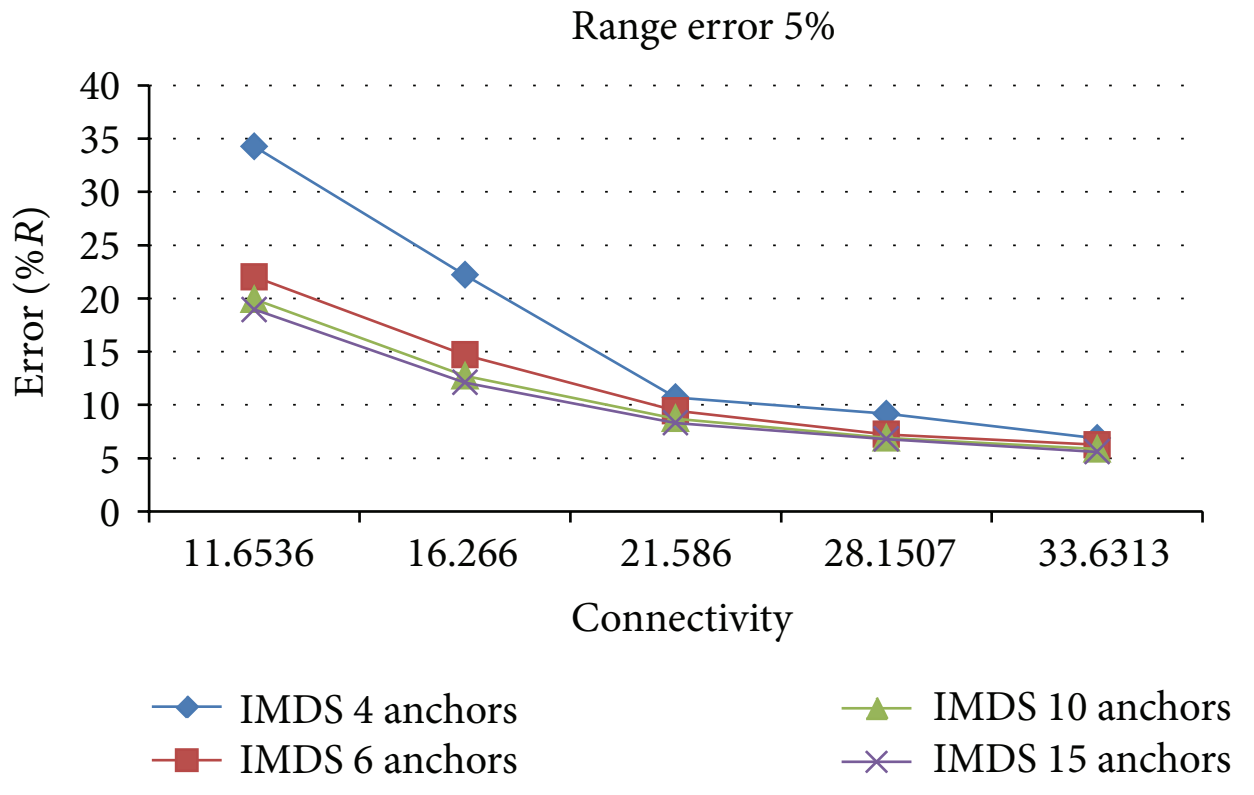

Figure 6: The effect of number of anchors on the estimation error.

## 6. Comparison of MDS-MAP and IMDS for 3D Surface WSN

Figure 8 shows an example of two typical 3D surfaces. On the upper picture, there is a surface which represents a valley, while the lower surface represents a mountain. In our simulations, two scenarios are constructed to emulate a terrain with a valley and a terrain with a mountain. 100 nodes are deployed randomly with a uniform distribution over these two surfaces.

Thus, 280 different networks were simulated and each node location was discovered with both MDS-MAP and IMDS technique.

It is expected that MDS-based algorithms for WSN localization will not work well for such scenarios, basically because of multihop distance between each pair of nodes. Our improved heuristic approach presented in this paper is expected to achieve more acceptable accuracy.

Figures 9 and 10 compare the results of MDS-MAP and IMDS for valley and mountain, respectively.

In the case of valley (Figure 9), when $e_r$ is small, both IMDS and MDS-MAP produce very similar estimation error. This error is much more affected by the number of anchors. As $e_r$ increases, IMDS performs much better than MDS-MAP for all connectivity levels, regardless of the number of anchors.

In case of mountain, for small $e_r$ MDS-MAP has smaller estimation error than IMDS (Figure 10). For large values of range error $e_r$, IMDS is better than MDS-MAP in terms of accuracy.

The average performance of IMDS as a function of connectivity for valley WSN is given in Figure 11. IMDS is very stable and predictive. Estimation error decreases as connectivity increases. The radio range error $e_r$ affects the estimation error in a way that larger $e_r$ deteriorates the performance of IMDS.

If we compare the results for valley and mountain, we can notice that both MDS-MAP and IMDS show better performance for valley terrain. The main reason for this is the characteristic of the terrain. Valley terrain is very regular because all nodes that are within radio range $R$ can communicate with each other. Mountain terrain should be considered as an irregular topology. The mountain presents an obstacle that obstructs the radio propagation between the nodes, which means that sometimes nodes that are very close to each other cannot communicate, that is, cannot measure the distance between each other. For terrains with obstacles, nodes localization problem should be solved differently. IMDS algorithm should manage hierarchical network organization based on cluster formation.

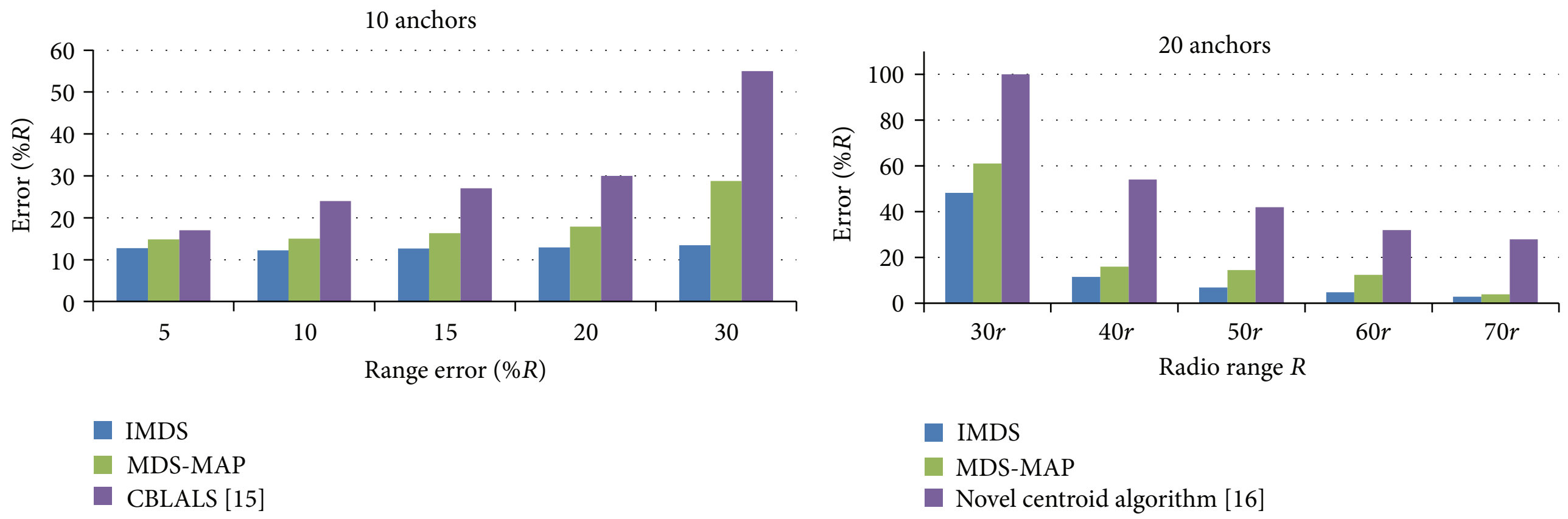

(a) (b)

FIGURE 7: Comparison of IMDS and MDS-MAP with CBLALS [15] and novel centroid algorithm [16].

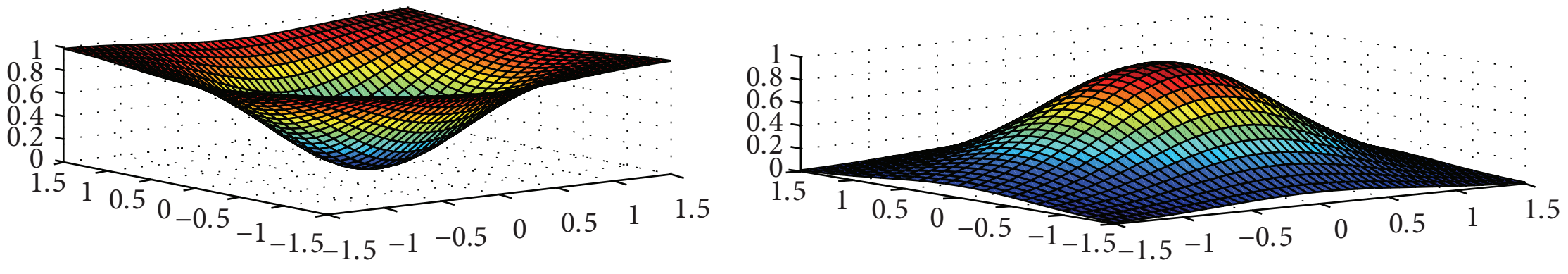

(a) (b)

FIGURE 8: Typical 3D surface, valley (a) and mountain (b).

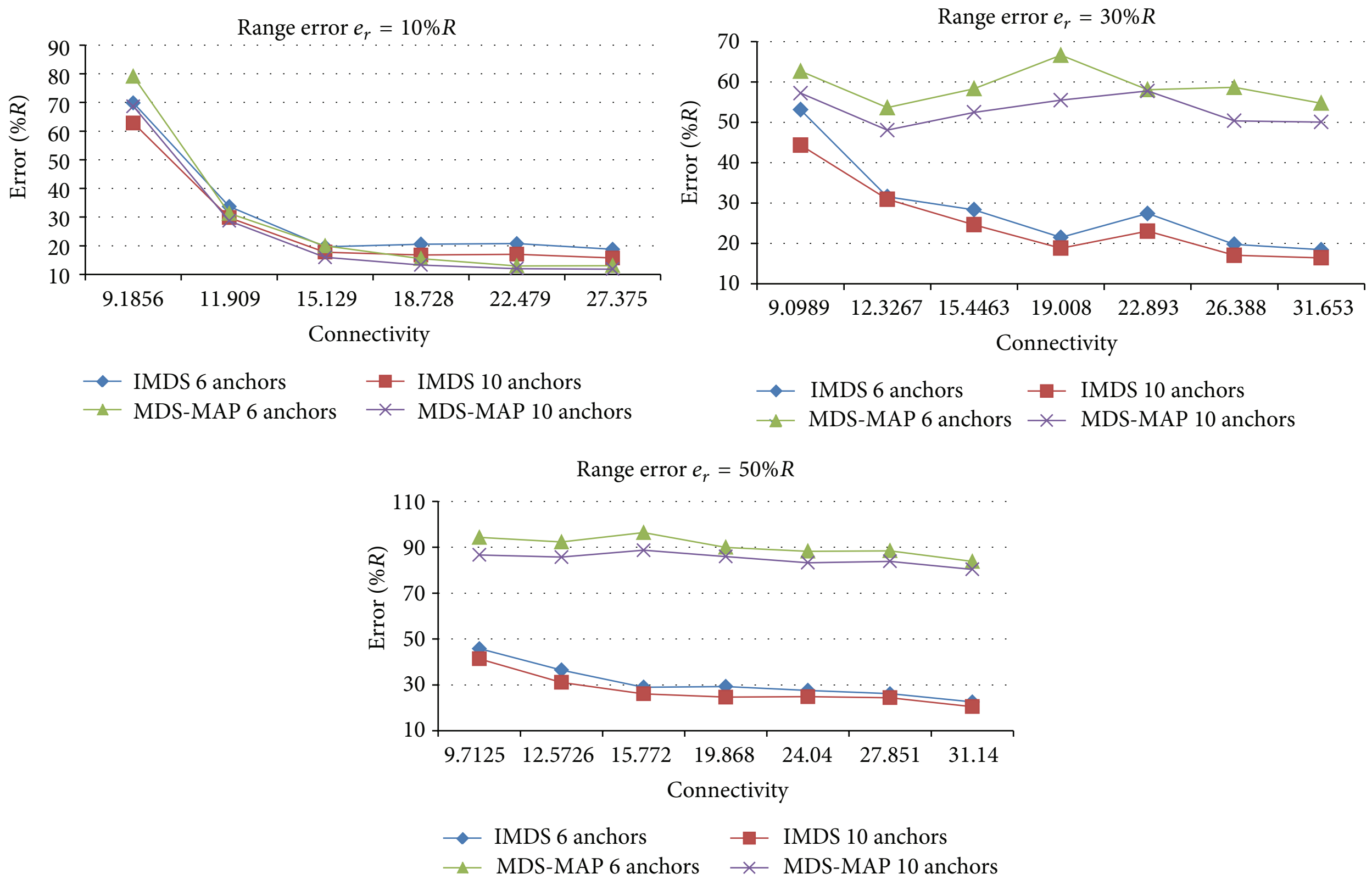

FIGURE 9: Comparison of MDS-MAP and IMDS for valley.

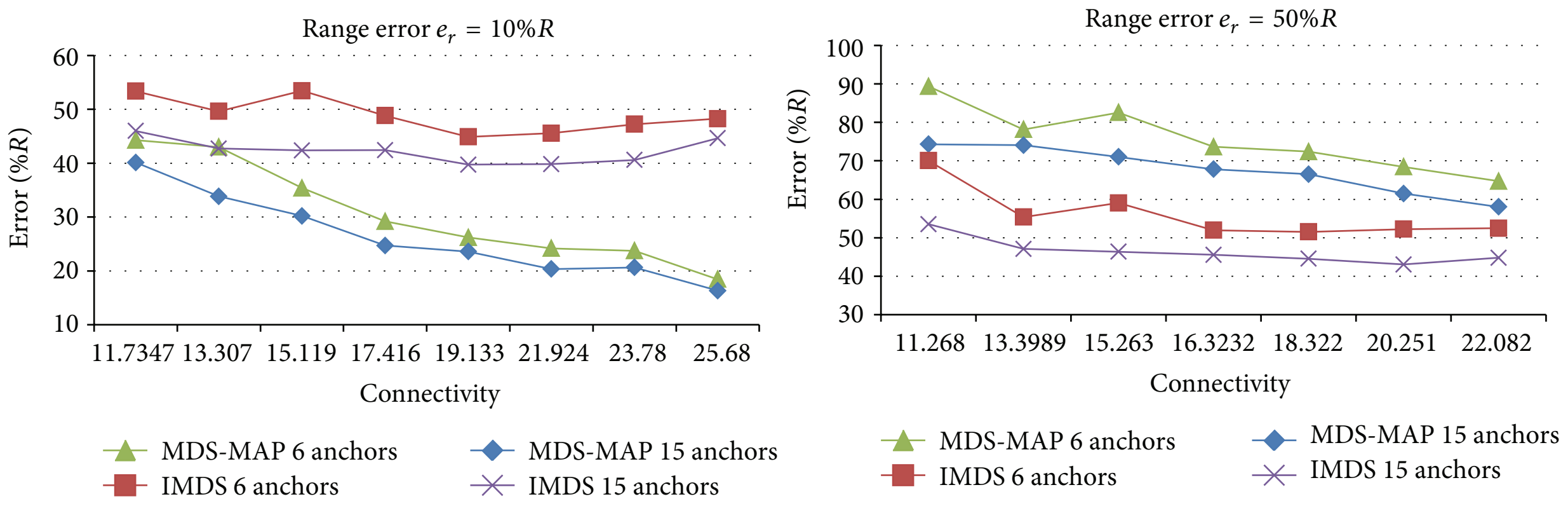

Figure 10: Comparison of MDS-MAP and IMDS for mountain.

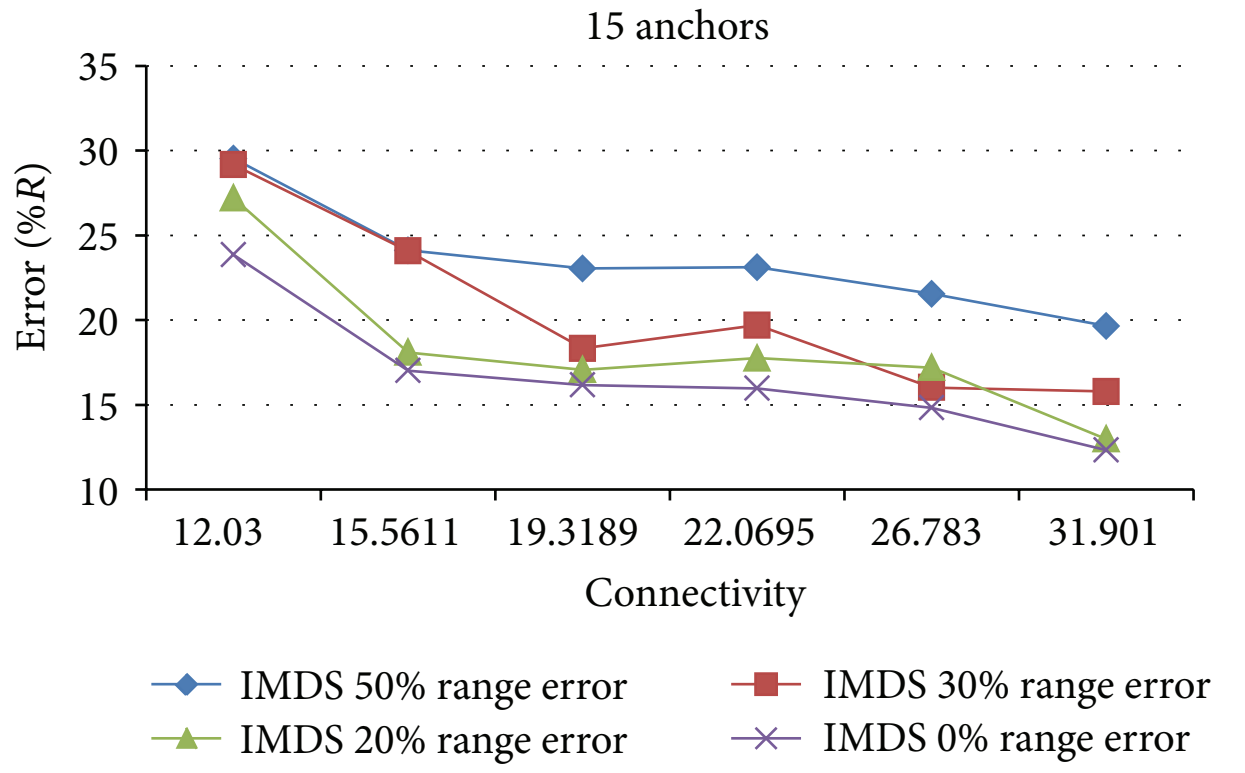

Figure 11: The effect of range error on the estimation error for valley.

## 7. Conclusion

In this paper, we investigated well known MDS-MAP algorithm for nodes localization in 3D-WSN. Additionally, we propose a new algorithm based on MDS that uses heuristic approach for distance matrix calculation (IMDS). Through extensive simulations, we evaluated the algorithms and showed that our IMDS algorithm outperforms other algorithms presented in [15, 16] in terms of accuracy.

For future work, we plan to investigate IMDS on irregular three-dimensional network topologies, where nodes are deployed on more complex 3D terrains. It is expected that MDS-based algorithms for WSN localization will not work well for such scenarios, basically because of multihop distance between each pair of nodes. For the future work we intend to extend IMDS considering hierarchical network organization based on cluster formation. This cluster-based approach which is already developed and implemented for 2D networks in [23] encourages us to consider cluster-based extension for 3D networks.

## Conflict of Interests

The author declares that there is no conflict of interests regarding the publication of this paper.